\documentclass[
footinbib,
a4paper,
aps,
prx,
superscriptaddress,
twocolumn,
preprintnumbers,
10pt]{revtex4-2}

\usepackage{graphicx}   
\usepackage{dcolumn}
\usepackage{mathrsfs}

\usepackage[utf8]{inputenc}
\usepackage{textgreek}

\usepackage{tikz}
\usetikzlibrary{quantikz2}
\tikzset{
  noisy/.style={starburst, line width=0.5pt, inner xsep=-3pt, inner ysep=-3pt}
}

\usepackage[colorlinks=true, citecolor=myblue, linkcolor=myred]{hyperref}

\usepackage{xcolor}
\definecolor{cmblue}{rgb}{0.12156862745098039, 0.4666666666666667, 0.7058823529411765}
\definecolor{mygrey}{gray}{0.35}
\definecolor{myblue}{rgb}{0.2,0.2,0.8}
\definecolor{mygreen}{rgb}{0.2,0.8,0.5}
\definecolor{myzard}{cmyk}{0,0,0.05,0}
\definecolor{mywhite}{rgb}{1,1,1}
\definecolor{myred}{rgb}{1,0.,0.3}
\definecolor{cyanZ}{RGB}{153,204,255}
\definecolor{redX}{RGB}{255,121,75}
\definecolor{pasqal}{RGB}{255, 152, 110}

\usepackage{dsfont}
\usepackage[normalem]{ulem}
\usepackage{bbold}
\usepackage{soul}
\usepackage{slashed}
\usepackage{enumitem,kantlipsum}

\usepackage{amsmath}
\usepackage{amsthm}
\usepackage{amsfonts}
\usepackage{amssymb}
\usepackage{amstext}
\usepackage{mathptmx}
\usepackage{bm}
\usepackage{xfrac}
\usepackage{mathrsfs}
\usepackage{latexsym}
\usepackage{array}

\usepackage{color}

\def\beq{\begin{equation}}
\def\eeq{\end{equation}}
\def\barray{\begin{eqnarray}}
\def\earray{\end{eqnarray}}
\makeatletter
\def\@seccntformat#1{\csname the#1\endcsname.\quad}
\makeatother

\usepackage[flushleft]{threeparttable}

\usepackage{newtxtext}
\usepackage{newtxmath}

\begin{document}

% Defining the title and authors
%\title{A Framework for Quantum Advantage through Quantum-Centric Supercomputing}
\title{A framework for quantum advantage}
\author{O. Lanes}
\affiliation{IBM Quantum, IBM Thomas J. Watson Research Center, USA}
\email{olivia.lanes@ibm.com}

\author{M. Beji}
\affiliation{PASQAL SAS, 24 rue Emile Baudot - 91120 Palaiseau, Paris, France}
\email{alexandre.dauphin@pasqal.com}

\author{A. D. C\'orcoles}
\affiliation{IBM Quantum, IBM Thomas J. Watson Research Center, USA}

\author{C. Dalyac}
\affiliation{PASQAL SAS, 24 rue Emile Baudot - 91120 Palaiseau, Paris, France}

\author{J. M. Gambetta}
\affiliation{IBM Quantum, IBM Thomas J. Watson Research Center, USA}

\author{L. Henriet}
\affiliation{PASQAL SAS, 24 rue Emile Baudot - 91120 Palaiseau, Paris, France}

\author{A. Javadi-Abhari}
\affiliation{IBM Quantum, IBM Thomas J. Watson Research Center, USA}

\author{A. Kandala}
\affiliation{IBM Quantum, IBM Thomas J. Watson Research Center, USA}

\author{A. Mezzacapo}
\affiliation{IBM Quantum, IBM Thomas J. Watson Research Center, USA}

\author{C. Porter}
\affiliation{IBM Quantum, IBM Thomas J. Watson Research Center, USA}

\author{S. Sheldon}
\affiliation{IBM Quantum, IBM Thomas J. Watson Research Center, USA}

\author{J. Watrous}
\affiliation{IBM Quantum, IBM Thomas J. Watson Research Center, USA}

\author{C. Zoufal}
\affiliation{IBM Quantum, IBM Research Europe – Zurich, 8803 R\"uschlikon, Switzerland}

\author{A. Dauphin}
\affiliation{PASQAL SAS, 24 rue Emile Baudot - 91120 Palaiseau, Paris, France}

\author{B. Peropadre}
\affiliation{IBM Quantum, IBM Research, Cambridge, MA 02142, USA}

\email{bperopadre@ibm.com}

% Adding abstract
\begin{abstract}

As quantum computers approach the threshold where they can demonstrably outpace their classical counterparts on certain tasks, the need for a precise, consensus-driven definition of quantum advantage becomes essential. Rapid progress in the field has blurred this term across organizations, architectures, and application domains. Here, we aim to articulate a functional definition for quantum advantage that is both platform-agnostic and empirically verifiable. Building on this framework, we highlight the algorithmic families most likely to achieve early advantage. Finally, we outline our vision for the near future, in which quantum computers enhance existing high-performance computing platforms.
\end{abstract}

% Generating the title
\maketitle

% Defining quantum advantage section
\section{Introduction}\label{sec:intro}
Quantum computing has long promised to solve computational problems that are intractable for classical computers~\cite{feynmann, shor, deutsch, Harrow2009, lloyd, grover, preskill}. Recent advances in quantum hardware, software, and algorithm design have significantly accelerated progress toward this goal. Notably, several experimental demonstrations have shown that quantum processors can now perform accurate computations at scales that surpass the capabilities of brute-force classical simulation of quantum circuits ~\cite{Kim2023, arXiv:2405.05068}. As these technologies continue to mature, a critical question has emerged at the forefront of the field: When will quantum computers deliver a clear, demonstrable advantage in solving practical, real-world computational tasks? This milestone, commonly referred to as {\it quantum advantage}, represents a pivotal threshold for the field. 

%It serves not only as a benchmark for quantum technological progress, but also as a guiding framework for allocating research investments and identifying application domains with near-term impact potential. Establishing criteria for when and how quantum advantage is achieved is therefore essential for advancing both our scientific understanding and the deployment of quantum technologies in real-world settings.

In this work, we propose a functional definition of quantum advantage and delineate the criteria required to substantiate such a claim. Establishing such criteria for when and how quantum advantage is achieved is crucial, not only as a rigorous benchmark of quantum technological progress, but also for identifying application domains with the potential for near-term impact. We also identify important components in this path: methods for the reliable execution of quantum circuits on performant quantum processors, their integration with high-performance computing (HPC) infrastructure, and the development of algorithms that can be executed with high fidelity and yield verifiably correct results.

\section{Definition of Quantum Advantage}\label{sec:definition}
Quantum advantage denotes the execution of an information processing task on quantum hardware that satisfies two essential criteria: i) the correctness of the output can be rigorously {\it validated}, and ii) it is performed with a {\it quantum separation} that demonstrably offers superior efficiency, cost-effectiveness, or accuracy than what is attainable with classical computation alone. Although the concept appears straightforward, these criteria remain insufficiently articulated in the current literature.%- e.g. in order to have confidence in the correctness of the output produced by a quantum computer, one must demonstrably validate that the quantum computation is trustworthy.

\subsection{Validation} Establishing trust in the outcome of a computation is one of the foremost tasks in establishing any new computational method. This is particularly critical for a quantum computation, since quantum computers are poised to solve problems that are not efficiently solvable by established classical methods, making a direct comparison challenging. Furthermore, since quantum computing is such a new paradigm, the concern is that simple sanity checks based on reasonable intuition may not be sufficient. Testing on simple cases that can be efficiently evaluated exactly by classical means may not capture the challenges the computation faces when addressing harder instances of the problem. As we argue in the manuscript, there are different ways in which we can ensure that the quantum computation can be validated to produce the correct answer even at a scale where establishing trust in the computation becomes challenging. The most straightforward way is to ensure that the computation itself was performed correctly by providing rigorous error bars, as can be obtained from a fault-tolerant quantum computer, or formally proven error mitigation methods. However, there are also interesting computational problems where the possibility of validation is directly possible by the nature of the answer.

\begin{enumerate}[wide, labelwidth=!, labelindent=0pt]
\item {\bf Error bars and error bounds} A universal way to ensure that the computation can be trusted is to ensure that every step of the computation was performed accurately. The way to achieve this formally is through proven error bars. There are different methods that can ensure such error bars, the cardinal approach being fault-tolerant quantum computing, cf. Section~\ref{sec:methods}1. Other methods, such as rigorously proven quantum error mitigation methods, cf. Section~\ref{sec:methods}2, and post-selected error detection,  cf.Section~\ref{sec:methods}3, are also able to provide rigorous error bars. However, compared with a fault-tolerant quantum computer, error mitigation and error detection methods have a cost that is more favorable only at smaller scales, cf. Section~\ref{sec:exp_values}, which is then clearly outperformed by fault-tolerant quantum computation at larger scales.
	
\item {\bf Problems with efficient classical verification} 
It is believed that for certain problems the answers can only be found efficiently on a quantum computer. But some of these problems are structured in such a way that, once found, these answers can be verified efficiently on a classical computer. Examples include sampling problems, cf. Section~\ref{sec:sampling},  where we are promised samples from a distribution with a specific structure, such as peakedness~\cite{arXiv:2404.14493}, or algebraic problems such as determining the prime factors of an integer~\cite{shor}. For these examples, we don’t need to rigorously check every step of the quantum computation, as we can directly verify the output is correct using a classical computer.

\item {\bf Variational problems} Similarly, there are problems for which it may not be possible to efficiently verify the solution on a classical computer directly, but we can easily score the quality of a solution and compare it with competing classical approaches. This is, for example, the case in variational problems, cf. Section~\ref{sec:sample_based}, such as in the estimation of the ground state energy of a molecule. In variational approaches, we are always assured that the reported number is an upper bound to the true energy value. This can happen without a statistical error~\cite{arXiv:2405.05068} or in presence of it~\cite{wu2024variational}.
Hence, solutions can be scored based on which one gives the lowest estimate. The same is true for classical optimization problems, where the solution with the smallest cost function value is the best answer. This can be established without ensuring that every step of the computation was performed accurately.   

\end{enumerate}
	
Although a formal validation of the computation is the foremost goal in establishing a quantum advantage, it is important to recognize the high relevance that experimental sanity checks play in building trust in the device. In particular, formally proven results always rely on a set of initial assumptions, such as accurate noise models for the quantum hardware or detailed knowledge of the underlying architecture, that must themselves be verified. These assumptions are usually easier to verify than the full quantum computation on its own. Additionally, heuristic methods~\cite{abanin2025constructive,filippov2023,Kim2023,haghshenas2025} play an important role in evaluating the capabilities of quantum computers. While these methods are not formally proven, they are valuable in two ways: first, they help lay the groundwork for a rigorous error analysis; and second, they provide experimental evidence for empirical quantum separations, paving the way for more concrete claims of quantum advantage.

\subsection{Quantum separations} 
Strictly speaking, it is not known whether a substantial, unconditional algorithmic separation between quantum computing and classical computing exists. Although this is considered extremely unlikely, it remains possible that the set of problems efficiently solvable by quantum computers could still be within the reach of classical methods. Unconditional separations have been proven in specific settings~\cite{bravyi2018quantum}, but these typically do not lead to the substantial, potentially exponential differences in performance that quantum computers are poised to deliver. All currently known separations depend either on the best known classical algorithms, as in the case of Shor's factoring algorithm~\cite{shor}, or on assumptions in complexity, as for example in ref.~\cite{arXiv:1612.05903}, cf. Section~\ref{sec:sampling}.  In practical terms, any claimed separation must withstand the test of time and be verified against both existing and emerging classical computational methods.

Therefore, once claims of an experimentally demonstrated quantum advantage begin to emerge, they must be evaluated on a common basis according to the criteria articulated in this article. Furthermore, quantum advantage is unlikely to materialize as a single, definitive milestone; instead, it should be understood as a sequence of increasingly robust demonstrations. Any claim of quantum advantage should be framed as a falsifiable scientific hypothesis, one that invites scrutiny, reproduction, and challenge from the broader research community. A cautious, iterative, and interpretive stance is therefore essential until state-of-the-art classical methods have had adequate opportunity to challenge the claim. However, should a classical approach later supersede a previously reported quantum advantage benchmark, the outcome should not be viewed as a failure; rather, it exemplifies constructive behavior that drives progress in both classical and quantum computation.

\section{Problems poised for quantum Advantage}\label{sec:problems}

In what follows, we discuss three key families of computational problems that currently stand out as promising avenues for realizing quantum advantage: i) problems that are solved via sampling algorithms, ii) problems that are verifiable via the variational principle, and iii) problems whose solution reduces to calculating expectation values of observables. We examine each of these approaches in greater detail, with the aim of elucidating their respective potential for realizing quantum advantage.

% Exploring sampling algorithms
\subsubsection{Sampling problems}\label{sec:sampling}

Sampling algorithms use all the information contained in a set of samples drawn from probability distributions produced by quantum circuits. Based on our current understanding of computational complexity theory, such algorithms are well positioned to address problems that are classically intractable — for example, integer factorization using Shor’s algorithm~\cite{shor}. In this specific case, it is straightforward to verify that a proposed solution
is correct through an efficient classical computation.  However, the practical realization of these algorithms is hindered by the need for fault-tolerant quantum computation, which remains significant challenges that are unlikely to be resolved in the near term.

Similarly, the quantum approximate optimization algorithm (QAOA) outputs candidate solutions for combinatorial problems~\cite{farhi2014quantum} that are classically verifiable. While we do not expect a general quantum speed-up for most optimization problems, we see potential for finding better solutions than classical solvers for {\it specific} problem instances with QAOA and other quantum heuristic approaches. To track progress towards potential quantum advantage and drive a fair and systematic benchmarking effort, a quantum optimization benchmark library has been introduced, with a curated set of 10 combinatorial optimization problem classes, corresponding problem instances, and well-defined reporting metrics~\cite{arXiv:2504.03832}. Crucially, for many problem classes, instances have already become challenging for state-of-the-art classical solvers at moderate problem sizes for known sparse model formulations.

Along similar lines, Cazals et al.~\cite{arXiv:2502.04291} have identified hard instances that can be natively suited for neutral-atom-based analog quantum computers. Furthermore, neutral atoms arranged in amorphous arrays have been proven to have a spin-glass phase~\cite{brodoloni2025spin}, which is expected to be hard to sample from by classical means. This indicates a strong compatibility with near-term quantum processors. Nevertheless, based on the comparative strength of classical optimization methods, we believe that the most promising path forward is to target individual problem classes and even specific problem instances.

Another prominent example within the sampling paradigm is random circuit sampling (RCS)~\cite{arXiv:1612.05903}, which has been proposed as a potential candidate to demonstrate quantum advantage. Yet, verifying the correctness of RCS outputs at scale is classically intractable. A natural approach to certifying such computations involves ensuring that the sampling process is effectively error-free. To date, the only universally accepted method for achieving this is fault-tolerant quantum computing. Consequently, we argue that several experimental claims of quantum advantage ~\cite{Nature.574.505,Nature.634.328,arXiv:2406.02501} remain unsubstantiated under this criterion. Alternative strategies for ensuring that individual samples are drawn faithfully include error detection techniques based on post-selection~\cite{fujii2016noise} and single-shot error mitigation using coherent Pauli checks~\cite{arXiv:2504.15725}, cf. Section~\ref{sec:methods}3. Exploratory efforts in analog quantum sampling — such as those employing neutral atoms~\cite{shaw2024benchmarking} or superconducting qubits in the XY model~\cite{andersen2025thermalization} — are promising from a physics perspective, but they similarly face challenges related to verification and robustness. 

In short, these demonstrations have not succeeded in certifying the accuracy of the underlying computations, thereby precluding a fair comparison between quantum and classical methods~\cite{kechedzhi2024effective}. Given the absence of efficient and scalable verification techniques, we conclude that random circuit sampling (RCS) does not yet constitute a fully satisfactory pathway to quantum advantage.

However, recent theoretical developments~\cite{arXiv:2404.14493} have introduced a promising variant of RCS based on peaked random circuits. These circuits are designed to generate a specific bitstring with non-negligible probability while preserving random-like characteristics across the remainder of the output distribution. This structure enables new opportunities for demonstrating verifiable quantum advantage, as the “peakedness” can be detected through an analysis of the sampled outputs. Although recent findings suggest that such peakedness may, in some cases, be simulated in quasi-polynomial time~\cite{arXiv:2309.08405}, this insight contributes to a deeper understanding of when and how quantum circuits exhibit classically tractable structure. Moreover, the possibility that classical algorithms might detect peakedness without requiring full simulation offers valuable guidance for designing more robust quantum advantage protocols. Overall, peaked random circuits provide fertile ground for exploring new complexity-theoretic frontiers towards verifiable quantum advantage, but will need more rigorous determination on hardness to be clear advantage candidates.

%However, recent theoretical work~\cite{arXiv:2404.14493} has introduced a promising variant of random circuit sampling (RCS) based on peaked random circuits. Such circuits are engineered to produce a specific bitstring with non-negligible probability, while maintaining random-like behavior otherwise. This structure opens new avenues for demonstrating verifiable quantum advantage, as the “peakedness” of the distribution can be detected by analyzing sampled outputs. Although recent results show that peakedness may, in some scenarios, be simulated in quasi-polynomial time~\cite{arXiv:2309.08405}, this insight also enhances our understanding of when and how quantum circuits exhibit a classically tractable structure. Furthermore, the possibility that classical algorithms might directly identify peakedness, without full simulation, offers valuable feedback for designing more robust quantum advantage protocols. Overall, peaked random circuits provide fertile ground for exploring new complexity-theoretic frontiers towards verifiable quantum advantage, but will need more rigorous determination on hardness to be clear advantage candidates.

% Discussing sample-based methods
\subsubsection{Variational %principle type of
problems}\label{sec:sample_based}

A promising family of computational problems for achieving quantum advantage consists of those governed by the variational principle. A key example is the estimation of ground-state energies in quantum systems—a foundational task underlying many challenges in condensed matter physics, high-energy physics, quantum chemistry, and materials science. The ground-state problem is particularly relevant in the context of quantum advantage for several reasons. First, when the algorithm employed satisfies the variational principle, approximate solutions can be classically ranked by their energy estimates, even in the absence of an exact solution. Consequently, if a quantum solution yields a solution with lower energy than any classically obtained counterpart, it offers a potentially verifiable instance of quantum advantage.

%A family of computational problems that are promising candidates for achieving quantum advantage is those that obey the variational principle. One such example is finding the ground-state energies in certain physical systems. Computing ground-state energies is an important application at the root of many computational challenges in condensed matter and high-energy physics, chemistry, and materials science. The ground-state problem has an important implication in the search for quantum advantage. First, if the algorithm used satisfies the variational principle, approximate solutions can be classically ranked by accuracy, according to the energy approximations they produce. As a consequence, if a quantum solution has a better accuracy or lower energy, that would indicate a possible verifiable quantum advantage. 

The degree of verifiability in quantum algorithms can vary significantly depending on the nature of the algorithm. One can have variational quantum solutions that come with/without an uncertainty error, or - for some algorithms - the entire solution proposed by the quantum computer can be classically representable. In the following sections, we examine several representative quantum algorithms, highlighting the spectrum of verifiability they offer and the implications for demonstrating quantum advantage.

%Verifiability can occur at different levels depending on the nature of the algorithm. One can have variational quantum solutions that come with an uncertainty error, without it, or - for some algorithms - the entire solution proposed by the quantum computer can be classically representable. In the following, we discuss a few examples of quantum algorithms that have a range of verifiability levels.

The Quantum Phase Estimation (QPE) algorithm~\cite{kitaev1995quantum} is one of the earliest quantum algorithms proposed for solving the ground-state energy problem. Given a suitable reference state with polynomial overlap with the true ground state, QPE can efficiently estimate the ground-state energy in polynomial time. The algorithm provides solutions with a precision error that decreases exponentially with the size of the quantum auxiliary register and its verifiability hinges on the trust that the quantum circuit is performed correctly and other approximations - such as those accompanying the Hamiltonian simulation part - are controlled. However, the quantum circuits that are necessary for the phase estimation algorithm are very deep and are generally thought to require fault-tolerant quantum computers.

%The Quantum Phase Estimation algorithm (QPE)~\cite{kitaev1995quantum} is one of the first algorithms proposed for the ground-state problem. Provided with a good reference state with a polynomial overlap with the true ground state, QPE can solve the ground state problem efficiently in polynomial time. The solution from QPE algorithm comes with a precision error which is exponentially small in the size of its quantum auxiliary register, and its verifiability hinges on the trust that the quantum circuit is performed correctly and other approximations - such as those accompanying the Hamiltonian simulation part - are controlled. However, the quantum circuits that are necessary for the phase estimation algorithm are very deep, and they are generally thought to require fault-tolerant quantum computers. 

In the last decade, variational quantum eigensolvers (VQEs)~\cite{peruzzo2014variational,kandala2017hardware} and their adaptive versions~\cite{grimsley2019adaptive} have been proposed as a low-depth alternative for ground-state problems, which can be executed on pre-fault-tolerant quantum devices. They can be seen as the quantum equivalent of variational quantum Monte Carlo methods~\cite{ceperley1977monte}. Compared to phase estimation, they do not have convergence guarantees, and are not directly tied to the quality of the initial reference state. Although the approximate estimations of the ground-state energies produced by VQE are variational, they come with an uncertainty related to the statistical nature of the energy estimation. This uncertainty allows for the possibility of approximate solutions to underestimate the true ground-state energy, complicating a strict ranking of the solution accuracy when compared to other methods. However, quality metrics that incorporate uncertainty, such as the V-score introduced in~\cite{wu2024variational}—have been proposed to better assess the quality of solutions. While still constituting a good option for lattice models, low-depth VQE is not a good choice for large-scale chemistry systems, since the number of measurements necessary to estimate molecular energy to useful precisions is prohibitive~\cite{wecker2015progress,zhang2025quantum}.  

Krylov Quantum Diagonalization (KQD)~\cite{arXiv:2407.14431} is a recently proposed algorithm that bridges the gap between QPE and VQE. Inspired by classical power methods, KQD inherits the convergence properties of QPE while also exhibiting the variational behavior and statistical uncertainties characteristic of VQE. Unlike VQE, it does not require the design of a variational ansatz, relying instead on quantum circuits that simulate real-time evolution. The verifiability of KQD solutions is comparable to that of VQE, allowing the solution accuracy to be ranked on the basis of energy estimates and associated uncertainties. However, like VQE, KQD faces significant limitations when applied to large-scale quantum chemistry problems due to the prohibitive number of measurements required to achieve high-precision energy estimates.

%The Krylov quantum diagonalization (KQD)~\cite{arXiv:2407.14431}, a quantum equivalent of classical algorithms based on power methods, bridges QPE and VQE, since it inherits convergence properties of QPE, and the variational behavior and statistical uncertainties of VQE. It removes the need to design a variational ansatz, and hinges on quantum circuits that simulate real-time evolutions. The verifiability of its solutions can be treated on the same level of VQE, and the solution accuracy can be ranked accordingly. Still, KQD cannot deal with large chemistry problems as a result of the same measurement problem that affects VQE.    

Recently, new quantum algorithms designed for execution on quantum-centric supercomputing platforms~\cite{arXiv:2405.05068, kanno2023qQSCI} have addressed several limitations of pre-fault-tolerant ground-state methods. Instead of estimating expectation values, as in VQE or KQD, these algorithms uses measurements from a quantum ansatz state to identify a subspace onto which the original Hamiltonian can be projected. The projected Hamiltonian has a much lower dimension than the original Hamiltonian and can be diagonalized using distributed classical computing. 
Remarkably, Sample-based Quantum Diagonalization (SQD) has already been used to estimate energies for systems that are beyond exact diagonalization, such as the iron sulfide cluster 2Fe-2S with results that match classical state-of-the-art approximate methods such as density matrix renormalization group, and has been used to tackle larger 4Fe-4S clusters~\cite{arXiv:2405.05068}. Since the final diagonalization is performed classically at machine precision, the resulting energy estimates are robust to quantum sampling errors, unlike VQE. Furthermore, the method mitigates the effect of errors in the quantum computer, removing the impact of useless quantum samples at the cost of classical overhead. 

The SQD approach can be further combined with KQD, by classically projecting a target Hamiltonian on a subspace of samples from real-time evolution circuits, in the sample-based Krylov quantum diagonalization (SKQD) workflow~\cite{arXiv:2501.09702}.
The benefit of incorporating SQD and KQD is twofold: first, it comes with convergence guarantees similar to phase estimation~\cite{Quantum.8.1457}, and second, it does not require the design of a quantum circuit ansatz.

Both the sample-based methods for ground states, SQD and SKQD, achieve the highest degree of verifiability. The fact that the approximate ground states produced by these methods can be stored in classical memory allows for a high-precision ranking of their accuracy, without the need to factor in uncertainties coming from statistical noise. Quantum advantage obtained with these methods would therefore be classically certifiable and reproducible. Quantum computers in the analog mode also allow one to generate bitstring distributions for regimes that can be challenging for classical methods. This includes regimes close to criticality or frustrated systems. Such examples are potential candidates for quantum-centric approaches such as SQD or SKQD.

% Calculating expectation values
\subsubsection{Expectation values of observables}\label{sec:exp_values}

The calculation of expectation values of observables from short-depth quantum circuits is a fundamental task across a range of near-term quantum algorithms. These include applications such as estimating molecular ground state energies~\cite{peruzzo2014variational}, evaluating quantum kernels in machine learning~\cite{havlivcek2019}, and simulating time-dependent spin correlation functions~\cite{Kim2023}. In the absence of fault-tolerant hardware, early implementations of variational algorithms demonstrated that such estimations are susceptible to noise-induced bias, even at modest circuit depths~\cite{kandala2017hardware}. The development of error mitigation techniques—discussed in detail in Section~\ref{sec:elements}—has provided a pathway to suppress~\cite{Temme2017, Kandala2019} and, in some cases, cancel~\cite{Temme2017} the effects of noise on observable estimation. While these methods often incur an exponential sampling overhead, recent improvements in hardware fidelity and sampling rates have enabled their application to quantum circuits with qubit and gate counts beyond the reach of brute-force classical simulation~\cite{Kim2023}. In several cases, the resulting estimates are now competitive with state-of-the-art classical approximations.

Simulating time evolution is a classic example of measuring expectation values of short-depth quantum circuits and a strong candidate for quantum advantage, that is important to consider from the lens of verifiability. Models with conserved quantities offer some verification of the quantum computation, but in the most general case, verifiability can be a challenge at scales beyond exact diagonalization. At such scales, as we will elaborate more in Section~\ref{sec:elements}, it is particularly crucial to have error mitigation methods that can provide expectation values with rigorous error bars, in order to place trust in the result of the quantum computation. 

%In this context, analog quantum computers have historically been the first to study time dynamics at scales beyond exact diagonalization~\cite{Choi2016}, albeit for restricted classes of models. Remarkably, it has been shown that quantum computers in the analog mode can yield reliable results even in the presence of errors, as physically relevant observables—particularly local and intensive quantities—remain stable under both deterministic and stochastic perturbations, with error bounds that are independent of the system size~\cite{trivedi2024quantum,arXiv:2311.14818,daley2022practical}. 

Simulating quantum time evolution is particularly well-suited to analog quantum simulators, which have already been used to explore many-body dynamics at scales beyond the reach of exact diagonalization for specific classes of Hamiltonians~\cite{Choi2016}. While the difficulty of achieving precise control and error characterization in these systems presents a general challenge for validating measured quantities, recent theoretical results suggest that certain physically meaningful observables—particularly local ones—can exhibit robustness to both deterministic and stochastic perturbations. Notably, the associated error bounds may remain independent of system size~\cite{trivedi2024quantum, arXiv:2311.14818, daley2022practical}.
These insights, combined with ongoing progress in analog error mitigation techniques~\cite{Guo2025, steckmann2025}, point to promising strategies for enhancing the reliability of analog quantum simulations.

%\\[2ex]

%Based on these algorithmic groups, several application domains are widely considered the most promising areas that could quickly take advantage of quantum advantage. Quantum simulations of chemistry and materials science~\cite{arXiv:2405.05068, georgescu2014quantum,Johnson2014, di2024quantum, alexeev2024quantum} are widely regarded as one of the most promising near-term applications. These problems are often intractable for classical computers due to the exponential complexity of simulating quantum systems with high accuracy. Quantum computers, in contrast, can efficiently represent other quantum systems, and quantum algorithms exist to simulate both static and dynamical properties of quantum systems that have no efficient classical counterpart and can, in principle, lead to quantum advantages. 
%On the classical application side, combinatorial optimization is a key target area of relevance in finance, logistics, scheduling, and other industrial applications. Although an exponential quantum advantage is generally not guaranteed for such problems, specific problem instances may benefit from heuristic quantum approaches that outperform classical heuristics in practice.

\section{Elements for Advantage}\label{sec:elements}

In what follows, we discuss crucial elements for the realization of a certifiable quantum advantage. These elements involve capabilities that enable the accurate execution of quantum circuits, with accuracy established through rigorous error bars. Finally, we discuss the role that quantum-centric supercomputing infrastructure, and performant quantum processors will play towards the realization of a quantum advantage. 

\subsection{Methods for running accurate circuits}\label{sec:methods}

\begin{enumerate}[wide, labelwidth=!, labelindent=0pt]

\item \textbf{Error correction}
%Quantum error correction~\cite{PhysRevA.52.R2493} (QEC) refers to a set of techniques that allow the protection of quantum information from decoherence, gate imperfections, and measurement errors — characteristics that are intrinsic to any realistic quantum computing architecture. In QEC, quantum information is encoded into logical qubits, i.e., a fault-tolerant representation of a quantum bit made of multiple physical qubits in such a way that errors affecting a subset of physical qubits can be detected and corrected without disturbing the encoded logical state. QEC lays the groundwork for fault-tolerant quantum computation, where quantum algorithms can be performed in a manner that is resilient to noise, provided that the physical error rates remain below a threshold value. This correctable nature of fault-tolerant circuits ensures the accuracy of quantum computations. However, given the substantial qubit and operational overhead required for encoding, syndrome extraction, and correction, the implementation of full QEC is not expected to be viable in the near term. As such, early quantum applications are likely to rely on other schemes such as error mitigation strategies.
Quantum error correction~\cite{PhysRevA.52.R2493} (QEC) refers to a set of techniques designed to protect quantum information from decoherence, gate imperfections, and measurement errors—limitations intrinsic to any practical quantum computing platform. In QEC, information is encoded into logical qubits: fault-tolerant constructs built from multiple physical qubits, such that errors on a subset of physical qubits can be detected and corrected without collapsing the encoded state.
QEC lays the groundwork for fault-tolerant quantum computation (FTQC), where quantum algorithms can be performed in a manner that is resilient to noise, provided that the physical error rates remain below a threshold. Among various QEC codes, the surface code~\cite{PhysRevA.86.032324} has been a leading candidate, for it is particularly suited for 2D architectures like superconducting qubits, and it offers high error thresholds close to 1\% per operation~\cite{Benito2025}. However, this code requires a large overhead—on the order of thousands of physical qubits per logical qubit for realistic noise models—posing a scalability challenge.
To address this, quantum low-density parity-check (qLDPC) codes have emerged as a compelling alternative. These codes can offer similar levels of error suppression with a 10x reduction in qubit overhead~\cite{Bravyi2024}. This makes them especially attractive for implementing deep-circuit algorithms such as Shor’s algorithm~\cite{shor} and quantum phase estimation (QPE)~\cite{kitaev1995}, which are potential candidates for achieving quantum advantage in areas such as integer factorization and quantum chemistry.

This correctable nature of fault-tolerant circuits ensures the accuracy of quantum computations. However, given the substantial qubit and operational overhead required for encoding, syndrome extraction, and correction, the implementation of full QEC is not expected to be viable in the near term, although recent algorithmic improvements could help change this paradigm~\cite{clinton2024}. As such, early demonstrations of quantum advantage will likely rely on other schemes such as error mitigation and error detection strategies.

\item \textbf{Error mitigation} Quantum error mitigation (QEM) encompasses a suite of techniques aimed at reducing—or in some cases, eliminating—the bias in expectation value estimates caused by noise in quantum circuits~\cite{Temme2017, Li2017}. Unlike quantum error correction (QEC), QEM operates through classical post-processing of noisy outputs from ensembles of circuit executions, without requiring additional qubits or fault-tolerant architectures. Common QEM strategies include readout error mitigation, which corrects for measurement assignment errors~\cite{nation2021scalable, Trex2022}, and zero-noise extrapolation (ZNE), where noise strength is deliberately amplified to enable extrapolation to the zero-noise limit~\cite{Kandala2019, giurgica2020digital, Kim2023}. Even in the absence of fault tolerance, several QEM methods have demonstrated the ability to yield accurate expectation values from short-depth circuits, with rigorous error bounds~\cite{PhysRevLett.119.180509, NaturePhys.19.1116}. However, these methods often incur significant sampling overhead, particularly due to the variance amplification introduced by quasi-probabilistic weighting. This is especially true for probabilistic error cancellation (PEC), which effectively inverts well-characterized noise channels~\cite{NaturePhys.19.1116}. The runtime cost of PEC scales approximately as $\sim (1 + 15\epsilon/8)^{nd}$ for an $n$-qubit circuit of depth $d$ and two-qubit depolarizing error rate $\epsilon$. Despite this exponential scaling, as hardware error rates $\epsilon$ continue to decline, QEM techniques can outperform classical methods even when the latter scale polynomially~\cite{arXiv:2407.12768} at the $\sim$100-qubit scale with thousands of two-qubit gates. We anticipate that continued improvements in hardware fidelity will further extend the reach of error mitigation to deeper circuits and more complex observables, bringing us closer to the realization of quantum advantage from expectation values~\cite{IBM_GammaBar, aharonov2025importance,zimboras2025myths}.

The reach of quantum error mitigation methods can be significantly extended through the integration of classical high-performance computing (HPC). By trading off additional classical runtime and tolerating controlled bias, classical resources can reduce the overhead associated with quantum error mitigation, enabling mitigation over larger circuit volumes at fixed hardware error rates. This synergy can leverage the latest advances in classical simulation. Notable examples include tensor network error mitigation (TEM)~\cite{filippov2023, arXiv:2411.00765}, which employs tensor networks to invert the noise channel; shaded light cone techniques for probabilistic error cancellation (PEC)~\cite{arXiv:2409.04401}, which limit the scope of noise inversion to causally relevant regions; and multi-product formulas (MPF) for algorithmic error mitigation in trotterized time evolution~\cite{Quantum.7.1067}, which improve accuracy without increasing circuit depth. These developments suggest that, even in the absence of full fault tolerance, quantum methods augmented by classical post-processing and simulation can soon surpass the capabilities of classical-only approaches. As hardware continues to improve and hybrid strategies mature, we anticipate that this convergence will play a pivotal role in the realization of quantum advantage.

\item \textbf{Error detection}
Post-selected quantum error detection methods represent a promising intermediate approach between full quantum error correction and classical error mitigation. Conceptually, one can view the distinction between error correction and error mitigation as a trade-off between quantum and classical resources—qubits versus sampling. Error detection occupies a middle ground: it uses additional qubits to detect, but not correct, errors. This includes schemes based on error-correcting codes with post-selection, which fall under the broader umbrella of error detection. 

Compared to error mitigation methods like probabilistic error cancellation (PEC), error detection can offer improved (though still exponential) sampling overhead, while requiring significantly fewer qubits than full error correction. Post-selected error detection is the process of detecting when a run of the circuit (a "shot") has been corrupted by an error, and post-selecting for good shots. The post-selection rate decreases exponentially with circuit volume and physical noise rate, but this approach benefits from extremely fast sampling speeds.

Another key advantage of error detection is its compatibility with single-shot measurements, rather than relying solely on expectation values. This allows quantum devices to produce higher-quality samples at the cost of reduced throughput—effectively slowing down the quantum computer to improve output fidelity. Moreover, error detection is inherently compatible with other error mitigation techniques, making it a flexible and powerful tool. The combination of better sampling overhead and its single-shot applicability makes error detection a promising candidate for achieving finite (but not asymptotic) quantum advantage.

Various methods have been proposed for error detection.  The most obvious approach is to use an error detection code such as the [[4, 2, 2]] code or the [[8, 3, 2]] code~\cite{PhysRevLett.122.080504, Gupta2024, Bluvstein2024, reichardt2024,reichardt2025, Self2024}. However, other alternatives exist for detecting errors not based on encoding of qubits but based on verifying symmetries in whole circuits~\cite{tsubouchi2025}. For example, we might expect that a particular circuit preserves particle number, and any measurement that reveals a violation of this expectation could indicate an error~\cite{PhysRevA.98.062339}. In recent years another approach based on Pauli checks has been introduced which can detect errors in Clifford-dominated circuits by non-destructively measuring Pauli operators around the circuit and verifying that the Paulis were correctly mapped to each other through the Clifford~\cite{PhysRevResearch.5.033193}. Each measurement acts as a check and detects some errors.

Since the process of error detection itself has extra overhead in the terms of qubits and gates, it can introduce new errors in the circuit. It is therefore crucial to design low-overhead error detecting circuits so that their net effect is the removal and not increase of errors. This was recently demonstrated in the framework of spacetime codes, with error detection in large graph states comprising up to 50 qubits and 2500 gates, and also creating the largest GHZ states on record of 100+ qubits~\cite{arXiv:2504.15725}.

\end{enumerate}

\subsection{Quantum-centric supercomputing infrastructure}\label{sec:qcsc}

A common misconception frames quantum advantage as a competition between quantum and classical machines. In practice, however, the relevant benchmark is not quantum systems in isolation, but hybrid quantum–classical systems that combine quantum processing units (QPUs) with classical supercomputers. As discussed in earlier sections, HPC can already play a critical role in extending the reach of quantum algorithms run on current pre-fault tolerant hardware —enhancing error mitigation and scaling ground-state simulations.

This emerging paradigm, which we refer to as quantum-centric supercomputing (QCSC), envisions QPUs as specialized co-processors tightly integrated with CPUs and GPUs. Classical processors remain essential throughout the quantum workflow, supporting tasks such as system control, calibration, circuit optimization, and post-processing~\cite{arXiv:2405.05068, arXiv:2411.00765, fuller2025improved}. The goal is not to replace classical computing, but to demonstrate that a hybrid architecture can outperform classical systems alone.

Realizing near-term quantum advantage—and preparing for fault-tolerant systems anticipated later in this decade—requires extending HPC infrastructure to support modular QPU integration. This vision is already taking shape: Pasqal and IBM quantum systems are being deployed in HPC centers~\cite{arXiv:2405.05068, Jones2025}, and both companies have released SLURM plugins to streamline QPU integration~\cite{sitdikov2025}.

These developments support a modular approach to algorithm design, where components can be independently benchmarked and optimized. Aligning quantum workloads with the parallelism of HPC programming models will be key. Workflow management tools—such as real-time sampling in variational algorithms like SQD—can further accelerate convergence and improve time-to-solution.

\subsection{Performant quantum processors}\label{sec:hw}
Achieving quantum advantage in the near term is fundamentally constrained by the performance of the underlying hardware. In particular, the number of qubits, the achievable algorithmic depth, the fidelity of qubit operations, and the efficiency of quantum–classical orchestration, collectively set the ceiling for any credible claim of advantage. Among the various quantum architectures under development, superconducting quantum processors~\cite{kjaergaard2020superconducting} and neutral atoms~\cite{RevModPhys.82.2313} stand out as particularly well positioned to reach this milestone, as both have recently surpassed the 100-qubit threshold and demonstrate promising progress across all key performance dimensions.
Other platforms, such as trapped-ion quantum processors~\cite{bruzewicz2019trapped}, also offer a compelling alternative path to quantum computation. These systems have been deployed in cloud-accessible environments and are already being integrated into early-stage hybrid quantum-classical workflows. While our focus in this work will remain on superconducting circuits and neutral atom systems, the broader landscape of quantum hardware continues to evolve rapidly, with multiple modalities contributing to progress toward quantum advantage.

%At IBM and Pasqal, we are developing quantum computers in this regime, specifically designed to support quantum-centric workflows—those that leverage quantum resources in close coordination with classical computing in a tightly orchestrated manner to address problems beyond the capabilities of classical methods alone. In what follows, we focus on two of the leading hardware paradigms that underpin current efforts towards advantage: superconducting circuits, as realized in IBM’s transmon-based architectures, and neutral atoms, as advanced by Pasqal’s scalable neutral atom platforms. These technologies represent distinct but complementary approaches to building quantum processors capable of supporting quantum-centric workflows and advancing toward quantum advantage. We detail their performance characteristics and integration with classical resources to illustrate how each contributes to the broader landscape of near-term quantum computing.

\subsubsection{Superconducting qubits} 
Superconducting quantum processors are among the most technologically advanced platforms to date~\cite{mckay2023,acharya2024quantum}. They offer high-fidelity, high-speed gate operations, modularity, and compatibility with integrated control electronics.

IBM’s quantum processors, for example, employ transmon qubits arranged in modular, scalable architectures~\cite{bravyi2022future, mckay2023}. The current 156-qubit Heron QPU achieves an %250+~$\mu$s $T_1$ relaxation times, 
error per layer gate (EPLG) of $2.3 \times 10^{-3}$~\cite{mckay2023}, %On the heron processor, at 5K gates and 100Q, 96$\%$ of weight-1 observables on a mirror circuit were within 10$\%$ accuracy using Probabilistic Error Amplification-based ZNE, 
and supports circuit layer operations per second (CLOPS) above 250,000~\cite{arXiv:2405.05068, qiskit_device_benchmarking_2025}.
Real-time classical controllers enable dynamic circuits and mid-circuit measurement and feedforward, allowing for error detection and mitigation protocols. The fast sampling rates of superconducting qubits makes them particularly attractive for near-term explorations of quantum advantage that rely on a large number of samples -- such as computations that rely on error mitigation or post-selected error detection capabilities, and algorithms like SQD. 

Finally, the hardware is ably supported by a comprehensive software stack, Qiskit, which provides a cloud-based runtime environment for low-latency quantum-circuit execution with classical feedback~\cite{Javadi2024}. This framework integrates seamlessly in a serverless environment, orchestrates distributed hybrid quantum–classical tasks across containerized back-ends, and provides tools for domain-specific compilation, optimization, and error mitigation routines. Together, these components reduce overhead in iterative algorithms and enables a high-performance development environment that extends the practical reach of near-term devices toward practical demonstrations of quantum advantage.

%Execution of hybrid quantum-classical workloads is supported by Qiskit, which provides a cloud-based runtime environment for low-latency quantum-circuit execution with classical feedback~\cite{Javadi2024}. This framework integrates seamlessly in a serverless environment, orchestrates distributed hybrid quantum–classical tasks across containerized back-ends, and provides tools for domain-specific compilation, optimization, and error mitigation routines. Together, these components reduce overhead in iterative algorithms and enables a high-performance development environment that extends the practical reach of near-term devices toward practical demonstrations of quantum advantage.

\subsubsection{Neutral atoms} 

Neutral-atom quantum computers are rapidly emerging as strong candidates for demonstrating quantum advantage, due to their scalability, flexible connectivity, and ability to operate in both analog and digital modes~\cite{henriet2020quantum,aquila2023}. These platforms trap individual atoms in programmable optical tweezers and control their interactions via Rydberg excitation, enabling the realization of both Ising-type analog dynamics and gate-based digital quantum circuits.
As a proof of this, Pasqal’s neutral-atom platform has already demonstrated 100+ qubits operating in analog mode and is expected to reach 250 qubits by 2026. In the analog regime, neutral atoms are naturally suited to simulate spin models, quantum dynamics and solve optimization problems, with demonstrated capabilities for addressing complex many-body systems. Pasqal QPUs are already being deployed alongside classical HPC resources at Tier-1 centers like GENCI and Jülich, where they can be accessed through standard schedulers and integrated into hybrid CPU/GPU/QPU workflows. 
Neutral atoms can also support digital, gate-based quantum computing, aiming at early demonstrations of fault tolerance. This provides a clear path from near-term application development to more robust, general-purpose capabilities.
\\

\section{Conclusion and future outlook}\label{sec:conclusion}
 
We expect that credible evidence of quantum advantage is likely to emerge in one of the areas highlighted here within the next two years, enabled by sustained coordination between the high-performance-computing and quantum-computing communities. 
%Once these claims begin to emerge, they must be assessed on a common basis according to the two criteria articulated in this article. Furthermore, quantum advantage is unlikely to materialize as a single, definitive milestone. Instead, it should be understood as a sequence of increasingly robust demonstrations. Any claim of quantum advantage should be framed as a falsifiable scientific hypothesis, one that invites scrutiny, reproduction, and challenge from the broader community. A cautious iterative and interpretive stance is therefore essential until state-of-the-art classical methods have had adequate opportunity to challenge the claim. However, should a classical approach later supersede a previously reported quantum advantage benchmark, the outcome does not signify failure; rather, it exemplifies constructive behavior that drives progress in both classical and quantum computation.
Finally, we envision that the following best practices will be necessary to underpin credible advantage claims: (i) definition of standardized benchmark problems jointly with classical experts to ensure relevance and fairness, (ii) publication of detailed methodologies and datasets, sufficient to enable reproducibility of the benchmarks, and (iii) maintenance of an open-access leaderboards to track evolving performance. These practices will help ensure that quantum advantage claims are not only credible, but also constructive for the broader computational sciences.
\newline
% Acknowledging contributions
\begin{acknowledgments}
The authors thank the IBM Quantum and Pasqal teams for fruitful discussions and insightful comments on the manuscript.
\end{acknowledgments}

% Adding bibliography

%\bibliographystyle{apsrev4-2}
%\bibliography{quantum_advantage}

% Ending the document
\end{document}